\documentclass{mem}
\usepackage{graphicx}

\begin{document}

\title{Investigating the Late Stages of Stellar Evolution with Long Period Variables from MACHO and 2MASS}

\author{Oliver J. Fraser\inst{1}, Suzanne L. Hawley\inst{1}, \and Kem H. Cook\inst{2}}
\institute{Department of Astronomy, University of Washington \\ Box 351580, Seattle WA, 98195-1580 \\ \email{fraser@astro.washington.edu, slh@astro.washington.edu}
\and
IGPP, Lawrence Livermore National Laboratory \\ MS L-413 \\ P.O. Box 808 \\ Livermore, CA 94550 \\ \email{kcook@llnl.gov}
}

\authorrunning{Fraser}
\titlerunning{Long Period Variables from MACHO and 2MASS}

\abstract{
We are re-analyzing the MACHO variable star database to explore the relationships between pulsation, evolution, and mass loss in evolved stars. We will analyze the multi-periodic properties of long period variable (LPV) stars, 50\% of which could not be assigned any period in the original analysis. Recent results show that the missing stars may be an important element in understanding the origin of the period-luminosity sequences observed in the LMC. Our goal is to characterize the morphology and periodic properties of these stars, and then use the stars' 2MASS colors along with theoretical isochrones to understand their mass loss and evolution. We will develop a luminosity-independent criteria to classify the LMC LPVs, which can then be applied to Galactic LPVs. This will enable a synthesis of knowledge between LPVs in the LMC and the well-studied Galactic examples.}

\maketitle{}

\section{Variable Stars in MACHO: Earlier Findings and Current Limitations}

\begin{figure*}[th!]
\includegraphics[angle=-90, width=0.9\textwidth]{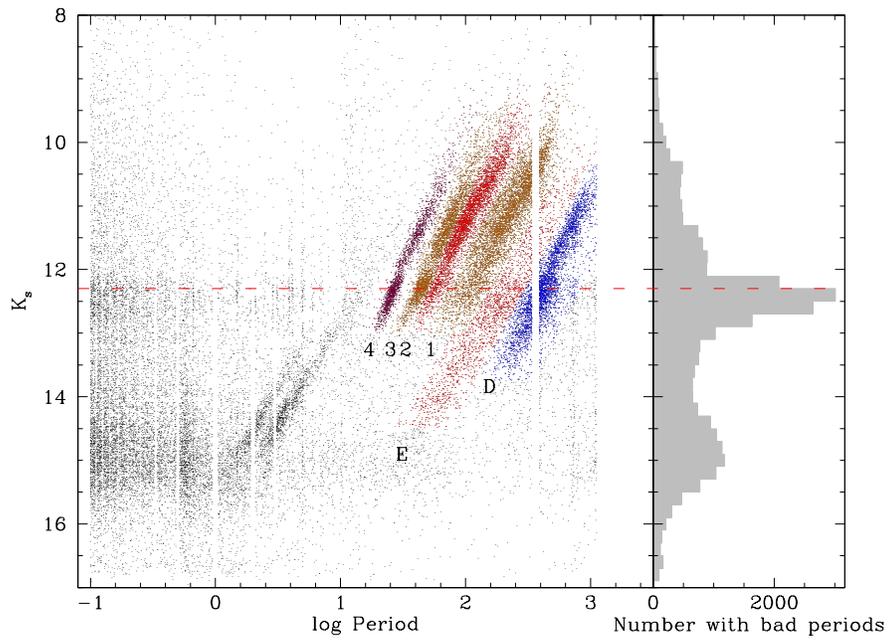}
\caption{\footnotesize LMC period-luminosity diagram showing the stars with well-determined periods. The histogram shows, for each luminosity bin, the number of stars with a poorly determined period. This includes 64\% of all the stars below the tip of the red giant branch.}
\label{perlum}
\end{figure*}

Long period variable stars (LPVs) in the Large Magellanic Clound (LMC) form a number of sequences in period-luminosity space (\cite{cook96, wood99}), as Figure \ref{perlum} (in 2MASS $K_s$) demonstrates. Each sequence has been well characterized by several authors, including our own group (\cite{fraser05}). However, only 50\% of the LPVs in the catalog can be shown here. The technique used to produce the MACHO variable star catalog was susceptible to aliasing or failing in cases where the star has no dominant period or pulsates with a very small amplitude. The majority (64\%) of stars with poorly determined periods lie below the tip of the red giant branch ($K_s = 12.3 \pm 0.1$ \cite{2000ApJ...542..804N}). Recent work shows that these stars may help explain some of the unexplained features in the period--luminosity diagram (\cite{Soszynski:2004wn}). One of our goals is to understand the characteristics of the pulsations of these stars, and how the patterns we find relate to the well-studied populations shown in Figure \ref{perlum}.

To better understand the stars with no simple period we are re-analyzing the MACHO variable star database using a date compensated discrete fourier transform algorithm. 

\section{Toward a Luminosity--Independent Classification of LPVs}

Microlensing surveys have revolutionized this field partly because their strategy of observing a large collection of stars at large distances lets us estimate the luminosities of these stars. Since we often do not know the luminosities of Galactic LPVs, there exists a dichotomy in LPV research between Magellanic Cloud stars and traditional Miras and semi-regular stars. One of our goals is to create a luminosity--independent classification for the LMC LPVs so that the same technique can be used to classify the light curves of Galactic stars. The immediate benefit of this work will be a synthesis in the knowledge from the Magellanic LPVs and the Galactic LPVs. We will also analyze other time-domain surveys to create Galactic catalogs of LPVs. In addition, our classification criteria will be more representative of the actual types of pulsation that are present in LPVs.

\vspace{1.5ex}
KHC's work was performed under the auspices of the U.S. Department of Energy, National Nuclear Security Administration by the University of California, Lawrence Livermore National Laboratory under contract No. W-7405-Eng-48.

\bibliography{../../refdb.bib}

\end{document}